\documentclass[twoside]{ilcws08}
\usepackage[latin1]{inputenc}
\usepackage[dvips]{graphicx,epsfig,color}
\usepackage{wrapfig,rotating}
\usepackage{amssymb,amsmath,array}

\begin{document}
\title{
%%%%   Paper title goes here  %%%%%%%%%%%%%%
Implementation of Particle Flow Algorithm and Muon Identification} %% 
%***********************************************************************
% AUTHORS INFORMATION AREA
%***********************************************************************
\author{M. J. Charles$^1$, U. Mallik$^1$ and T. J. Kim$^1$
% DO NOT MODIFY THE FOLLOWING '\vspace' ARGUMENT
\vspace{.3cm}\\
% Addresses and institutions (remove "1- " in case of a single institution)
1- Department of Physics and Astronomy, University of Iowa,\\
203 Van Allen Hall, Iowa City, IA 52242-1479, USA.\\
%% Remove the next three lines in case of a single institution
}
%%***********************************************************************
% END OF AUTHORS INFORMATION AREA
%***********************************************************************
\maketitle

\begin{abstract}
We present the implementation of the Particle Flow Algorithm and 
the result of the muon identification 
developed at the University of Iowa. We use Monte Carlo samples generated 
for the benchmark LOI process with the Silicon Detector design at the 
International Linear Collider.  
With the muon identification, an improved jet energy
resolution, good muon efficiency and purity are achieved.
\end{abstract}

\section{Introduction}
The majority of the interesting physics processes at 
the International Linear Collider (ILC)
involve multi-jet final states originating from hadronic decays of heavy gauge bosons. 
To fulfill its physics goals, the reconstructed di-jet mass resolutions for
$W$ and $Z$ decays should be comparable to their widths.
This di-jet mass resolution, in turn, requires di-jet energy resolution to
be as good as  
$3\%$\footnote{$\Delta M/M=\Delta E/E$ within the approximation $M = \sqrt{2E_{1}E_{2}(1-\cos\theta)}$}.
In order to achieve this goal,
the Particle Flow Algorithm (PFA) approach has been adopted by the Silicon Detector (SiD)
and the International Large Detector (ILD): two leading detector concepts at the ILC.

The goal of a Particle Flow Algorithm is to reconstruct events at the
level of individual particles, identifying charged and neutral particles 
separately in the calorimeters with minimal confusion.
The energy of the shower from each charged particle is replaced by the
momentum of its track measured in the tracking system which has orders  
of magnitude better precision than the calorimeters while 
the energy of photons and neutral hadrons are obtained from the calorimeters.
Without any confusion between charged and neutral showers in the calorimeters,
a theoretical jet energy resolution of approx. $\sigma_{\rm{E}} / \rm{E} = 20\% /\sqrt{\rm{E(GeV)}}$ can be reached.
In reality, however, it is not feasible to disentangle all charged showers from close-by neutral showers~\cite{ron}.
The shower leakage also has been shown as the another main source of deterioration of resolution~\cite{mat}. 
Due to confusion and leakage, the theoretical limit can not be reached.
Muon identification can reduce the confusion as presented in this paper.

We present here the current status of the PFA based on the SiD concept 
developed by the University of Iowa group in collaboration with SLAC National
accelerator Center (SLAC) and others.

\section{Particle Flow Algorithm at University of Iowa}
For the PFA performance, we use Monte Carlo (MC) samples of
$e^+e^- \rightarrow Z(\nu\bar{\nu})Z(q\bar{q})$, where $q=u,d,s$ at 500 GeV
and $e^+e^- \rightarrow q\bar{q}$ samples at 100, 200, 360 and 500 GeV 
with 10000 events each generated with GEANT4 simulation of SiD. 
LCIO serves as persistent data format. 

Initially the PFA at the University of Iowa was developed with
a `cheat' tracking. Now we use the tracking package~\cite{trk}
for the PFA while the cheat tracks can still be used for development.
Initial track and cluster association in PFA 
identifies photons, electrons and muons.
Electromagnetic (EM) showers are firstly reconstructed by looking at 
the shape and location of the cluster.
Clusters in the EM showers which are unassociated with
tracks are reconstructed as photons.
Clusters identified as the EM showers whose position coincides with a charged track and
whose energy matches the track's momentum are identified as electrons.
A muon is reconstructed
by matching the extrapolated track in the calorimeters, consistent with
energy deposited by a minimum ionizing particle (mip), with the muon direction
obtained from hits in the muon detector. 
The details are discussed in Section~\ref{sec:muon}.
These reconstructed tracks and clusters do not participate in
the later clustering for charged hadron showers.

Initial clustering is done for each sub-detector using hits
in the calorimeters and the muon endcap system.
Better jet energy resolution is obtained when the muon endcap
is used as a tail catcher~\cite{mat}. 
The barrel section is not currently used as a tail catcher 
because of the meter-thick solenoid between the hadronic calorimeter and the muon system:
this makes matching incoming tracks to hadronic showers correctly much more difficult.

Tracks found in the tracking system are extrapolated
as helices to the inner surface of the EM calorimeters.
Every track is supposed to have a seed
where the seed is defined as the `earliest' cluster
in the calorimeters directly connected to the extrapolated track.
When the track has too low momentum to reach EM calorimeters,
there is no seed found. 
In this case, the track in the  tracking system is used without clustering in the calorimeters.
When the seed of the track is not found due to early decay in the tracking system
even though the track reaches EM calorimeters,
the track is not used to avoid the double counting the energy 
assuming it deposited its energy somewhere in the calorimeters. 
If two tracks are overlapping, it is possible to have the same seed 
or connected via the showers.
In this case, the two or more tracks are grouped together 
to compare the energy deposited in the calorimeters with 
the momenta of these tracks.

The next important step of PFA is the 
assignment of the calorimeter hits correctly to the charged hadrons.
This also means the
efficient discrimination of showers produced by nearby charged
and neutral hadrons. We use a scoring system to add clusters to the shower.
A score is defined by the shower-related geometric quantities 
in the range from 0 to 1.
We assign the score to the link between two clusters based on how close two clusters are. 
Having assigned a score to the link, we build a 
charged hadron shower starting from seed of the track. The charged
hadron clusters are found and assigned to the shower of the track 
by searching the highest score link until 
the energy of clusters found are equal to the momentum of track within a
given uncertainty.
We iterate this process by loosening the selection criteria for clusters.

Even though electromagnetic showers are well-contained, consisting of a dense, almost
needle-like longitudinal core, it has a halo of nearby hits and back-scattering hits for 
low energy showers. 
Hadronic showers have clear internal structure and often
produce secondary neutral particles which deposit energy far from the main cluster. 
We take into account these back-scattering and secondary neutral particles 
by assigning the clusters in a conical order 
to the tracks whose energy in the calorimeters do not match the momentum of track 
measured by the tracking system.
The neutral hadron showers are from the remaining clusters not assigned to any track.

The last step of PFA is to reconstruct the four-momentum of all visible particles in an event.
The four-momentum of charged particles are measured in the tracking system while
the energy of photons and neutral hadrons are obtained from the calorimeters assuming
that the charged and neutral hadron are pion and kaon, respectively.
All reconstructed particles are filled into a reconstructed particle collection assigned to the event.

\section{\label{sec:muon}Muon Identification}
Muon identification performance has been measured with the
benchmark LOI samples like
$e^+e^- \rightarrow t\bar{t}$, $e^+e^- \rightarrow ZH$, $e^+e^- \rightarrow \tau\bar{\tau}$ 
and Standard Model (SM) background events~\cite{nor}.
These have plenty of muons and are used for physics benchmark analyses.

Muon identification procedure is done at the first stage in PFA. 
When clustering for charged hadron showers, this muon identification can reduce the confusion 
with charged hadrons by subtracting muon tracks and clusters at the beginning.
The minimum momentum to reach endcap and barrel muon
detector is 2 GeV and 5 GeV, respectively.
The muon identification study in this paper has been performed with  
tracks higher than 5 GeV, to reach either muon detector. 
Below 5 GeV, muons are badly reconstructed so the efficiency and purity are poor.

\begin{figure}
\begin{tabular}{cc}
{\includegraphics[width=0.45\columnwidth]{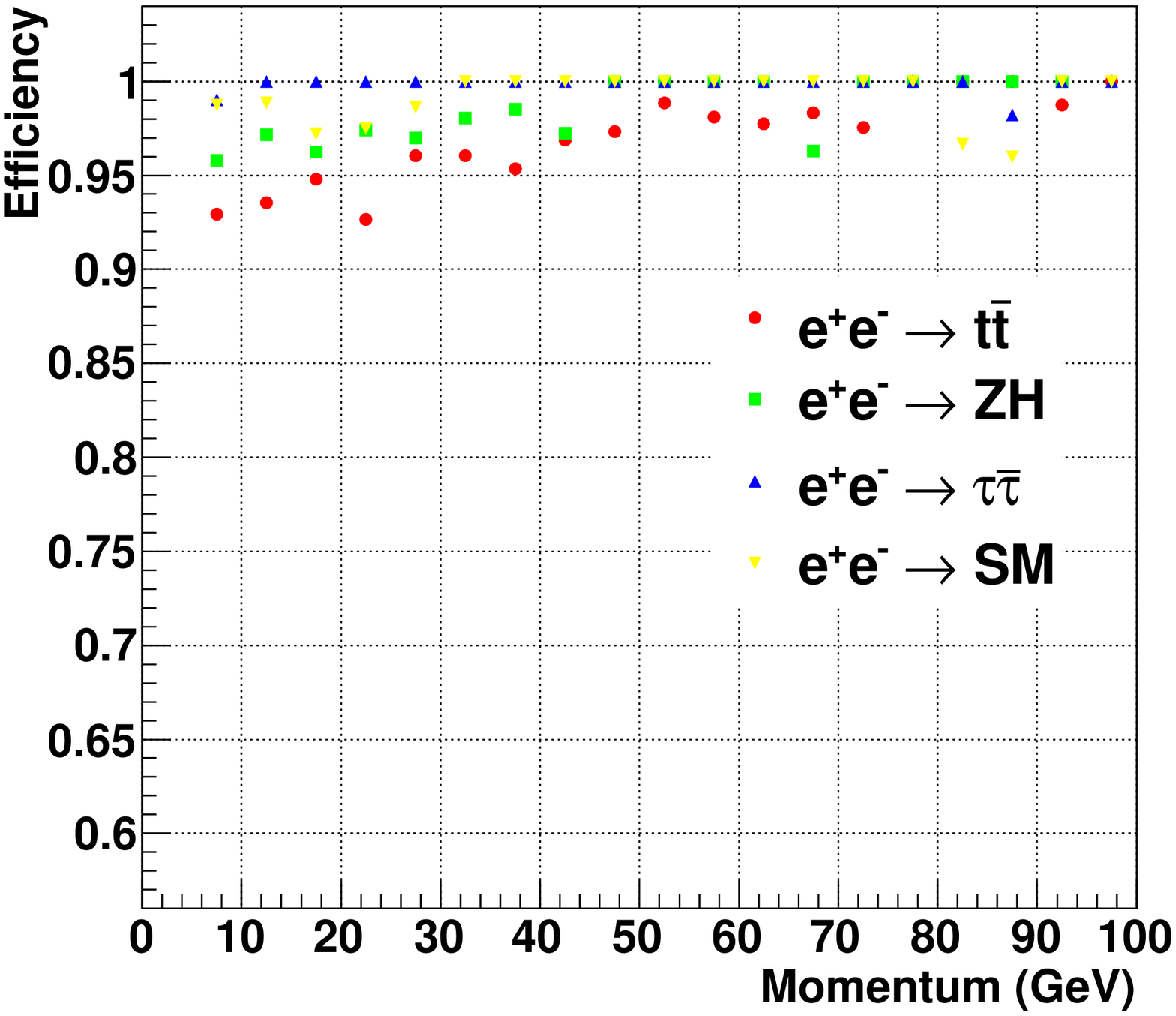}}&
{\includegraphics[width=0.45\columnwidth]{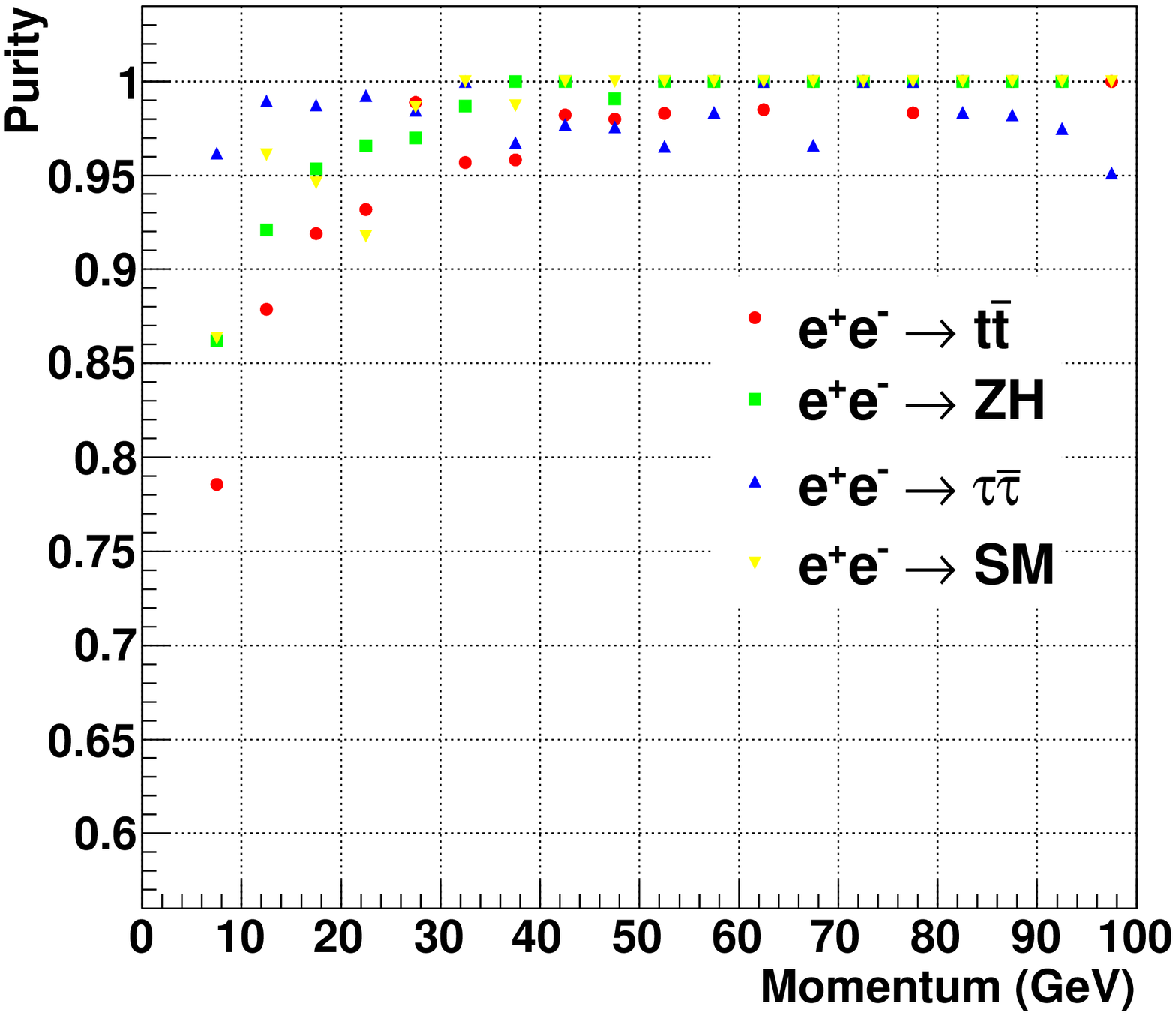}}\\
\end{tabular}
\caption{Efficiency (left) and purity (right) as a function of momentum for different samples:
$e^+e^- \rightarrow t\bar{t}$, $e^+e^- \rightarrow ZH$, $e^+e^- \rightarrow \tau\bar{\tau}$ and SM background events.}\label{fig:mu} 
\end{figure}

The muon identification algorithm starts with identifying the standalone muon mip 
in the muon detector. The standalone muon mip is defined 
as the group of isolated hits in the muon detector close to projective line
from the interaction point. 
The requirement of at least 5 isolated hits is optimized for high purity. 
Since SiD has a longitudinal segmentation of 20 cm thick iron return yokes
which is longer than transverse granularity of 3 cm~\cite{sid}, 
it is conceivable that the direction of mip in the muon detector can be obtained 
by using the first and second hit of standalone muon mip in the muon detector assuming 
that there is no magnetic effect outside the calorimeters so track's
trajectory is close to a straight line.
Track in the tracking system is extrapolated through 
the end of hadronic calorimeters using a helical trajectory used 
in simulation studies.
The tangent direction of the extrapolated track at the last layer in hadronic calorimeters
is compared with  
the direction of mip in the muon detector. Then, the best matched track is selected 
as the muon track. 

Figure~\ref{fig:mu} shows the muon efficiency and purity   
with different samples.
Efficiency is defined as the ratio of the number of the good reconstructed muons
to the number of real muon tracks using MC truth information.
Purity is defined as the ratio of the number of the good reconstructed muons to the number of reconstructed muons.
The average muon efficiency above 5 GeV is obtained
as 96\% for $t\bar{t}$, 97 \% for $ZH$, 100 \% for $\tau\bar{\tau}$ and 99 \% for SM background sample,
while the average muon purity above 5 GeV is also obtained
as 92\% for $t\bar{t}$, 93 \% for $ZH$, 98 \% for $\tau\bar{\tau}$ and 95 \% for SM background sample.

Purity worsens with deceasing track momentum. 
The stray magnetic field outside the calorimeters has an effect on the lower
momentum tracks even though it is not taken into consideration, leading to
inferior matching.
An additional source of misidentification as muons
is from weak decays of charged pion and kaons in the final state. 
Even though these are not true prompt muons, for the purpose of calorimeter 
pattern-recognition we exclude these tracks with its associated clusters 
from hadron shower clustering
as the mip-like behavior leave no showering in the calorimeters and muon systems. 

\section{Result}
\begin{table}[h]
\begin{center}
\begin{tabular}{l|cc|cc}
\hline
\hline
Sample    &  \multicolumn{2}{c|}{Barrel ($0 < \cos\theta < 0.8$)}  & \multicolumn{2}{c}{Endcap ($0.8 < \cos\theta < 0.95$)} \\ 
                 & Before  & After     &  Before   & After   \\ \hline
$e^+e^- \rightarrow q\bar{q},~\sqrt{s} = 100$ GeV    & 3.7 \%  & 3.6 \%   &   3.8 \%  & 3.6 \%  \\
$e^+e^- \rightarrow q\bar{q},~\sqrt{s} = 200$ GeV    & 3.0 \%  & 2.9 \%   &   3.2 \%  & 3.1 \%  \\
$e^+e^- \rightarrow q\bar{q},~\sqrt{s} = 360$ GeV    & 2.7 \%  & 2.6 \%   &   2.7 \%  & 2.6 \%  \\
$e^+e^- \rightarrow q\bar{q},~\sqrt{s} = 500$ GeV    & 3.5 \%  & 3.4 \%   &   3.3 \%  & 3.2 \%  \\
$e^+e^- \rightarrow Z(\nu\bar{\nu})Z(q\bar{q)}$   & 4.7 \%  & 4.7 \% &  3.9 \% & 3.8 \% \\\hline\hline
\end{tabular}
\end{center}
\caption{Energy resolution for di-jet samples of $\sqrt{s}$ = 100, 200, 360 and 500 GeV and mass resolution for $e^+e^- \rightarrow Z(\nu\bar{\nu})Z(q\bar{q})$ sample, where $q=u,d,s$ 
at $\sqrt{s}$ = 500 GeV in barrel and endcap region before and after muon identification. 
Both jets should be in the required angular region. $\cos\theta$ is defined as $|\frac{P_{z}}{P}|$.}
\label{table:resolution}
\end{table}

Table~\ref{table:resolution} shows the PFA performance is slightly better after muon identification.
In this table, the performance of the reconstruction is measured by the root-mean-square of 
the smallest range of reconstructed energies containing 90 \% of the events (rms$_{90}$).
This shows that muon identification improves the jet energy resolution overall energy range
in the reasonably good muon reconstruction efficiency and purity for above 5 GeV.
In the future, the optimization of each parameter in the algorithm 
is necessary to improve the resolution. 

% ****************************************************************************
% BIBLIOGRAPHY AREA
% ****************************************************************************

\begin{footnotesize}
% IF YOU DO NOT USE BIBTEX, USE THE FOLLOWING SAMPLE SCHEME FOR THE REFERENCES
% ----------------------------------------------------------------------------

% ----------------------------------------------------------------------------

\end{footnotesize}
\end{document}